\documentclass[aps,prb,twocolumn,floatfix,footinbib,superscriptaddress]{revtex4}
\usepackage[english]{babel}
\usepackage[utf8]{inputenc}
\usepackage[dvips]{graphicx}
\usepackage{fancyhdr}
\usepackage[active]{srcltx}
\usepackage{setspace}
\usepackage{float}
\usepackage{amsmath}
\usepackage{amsfonts}
\usepackage[]{natbib}
\usepackage{array}
\usepackage{color}
\usepackage{multirow}

\begin{document}

\title{Common nonlinear features and spin-orbit coupling effects in the Zeeman splitting of novel wurtzite materials}

\author{Paulo E. Faria~Junior}
\email{fariajunior.pe@gmail.com}
\affiliation{Institute for Theoretical Physics, University of Regensburg, 93040 Regensburg, Germany}

\author{Davide Tedeschi}
\affiliation{Dipartimento di Fisica, Sapienza Universit\`a di Roma, 00185 Roma, Italy}

\author{Marta De Luca}
\affiliation{Dipartimento di Fisica, Sapienza Universit\`a di Roma, 00185 Roma, Italy}
\affiliation{Department of Physics, University of Basel, 4056 Basel, Switzerland}

\author{Benedikt Scharf}
\affiliation{Institute for Theoretical Physics, University of Regensburg, 93040 Regensburg, Germany}
\affiliation{Institute for Theoretical Physics and Astrophysics, University of W\"urzburg, 97074 W\"urzburg, Germany}

\author{Antonio Polimeni}
\affiliation{Dipartimento di Fisica, Sapienza Universit\`a di Roma, 00185 Roma, Italy}

\author{Jaroslav Fabian}
\affiliation{Institute for Theoretical Physics, University of Regensburg, 93040 Regensburg, Germany}


\begin{abstract}

The response of semiconductor materials to external magnetic fields is a reliable 
approach to probe intrinsic electronic and spin-dependent properties. In this study, 
we investigate the common Zeeman splitting features of novel wurtzite materials, 
namely InP, InAs, and GaAs. We present values for the effective g-factors of different 
energy bands and show that spin-orbit coupling effects, responsible for the spin 
splittings, also have noticeable contributions to the g-factors. Within the Landau 
level picture, we show that the nonlinear Zeeman splitting recently explained in 
magneto photoluminescence experiments for InP nanowires by Tedeschi et al. [arXiv:1811.04922 (2018)] 
are also present in InAs, GaAs and even in the conventional GaN. Such nonlinear 
features stem from the peculiar coupling of the A and B valence bands, as a consequence 
of the interplay between the wurtzite crystal symmetry and the breaking of time-reversal 
symmetry by the external magnetic field. Moreover, we develop an analytical model to 
describe the experimental nonlinear Zeeman splitting and apply it to InP and GaAs data. 
Extrapolating our fitted results, we found that the Zeeman splitting of InP reaches 
a maximum value, which is a prediction that could be probed at higher magnetic fields.
 
\end{abstract}


\maketitle


\section{Introduction}
\label{sec:intro}

Novel III-V semiconductor compounds with wurtzite (WZ) crystal structure, such as 
InP\cite{DeLuca2014NL, Tedeschi2016NL, DeLuca2017APR}, 
InAs\cite{Moller2012Nanotech, Rota2016NL} 
and GaAs\cite{Furthmeier2016NatComm, DeLuca2017NL}, can nowadays be synthesized 
as nanowhiskers or nanowires (NWs)\cite{Dubrovskii2009Semiconductors, 
Carofff2009NatNano} with large diameters. In contrast to the widely studied zinc-blende (ZB) phase\cite{Vurgaftman2001JAP} --
the most stable crystal structure of non-nitride III-V compounds -- there are still 
many unknown, or at least not completely understood, properties of these WZ materials, 
especially regarding spin-dependent phenomena\cite{Zutic2004RMP}. For instance, 
spin-orbit coupling (SOC) parameters and effective g-factors in WZ NWs control 
the physics behind the exotic Majorana bound states in semiconductor/superconductor 
setups\cite{Lutchyn2010PRL,Oreg2010PRL,Das2012NatPhys,Albrecht2016Nature}, spin-laser operation\cite{Chen2014NatNano,FariaJunior2017PRB}, 
spin-relaxation mechanisms\cite{Kammermeier2018PRB,Dirnberger2018} and can be 
drastically modified under lateral quantum confinement\cite{Winkler2017PRL,Campos2018PRB}.

One of the possibilities to probe the intrinsic spin properties of a semiconductor 
system is to investigate their response under external magnetic fields, for instance, 
coupled to optical excitation in magneto photoluminescence (PL) experiments. 
Recent studies investigated the Zeeman splitting (ZS) from magneto 
PL\cite{DeLuca2013ACSNano,DeLuca2014NL,DeLuca2017NL,Tedeschi2018} and extracted 
effective g-factors using the conventional linear dispersion of the WZ ZS\cite{Cho1976PRB, Venghaus1977PRB}. 
Despite the successful description of the ZS for the magnetic field oriented 
{\it perpendicular} to the NW axis (with [0001] growth direction), this theoretical 
modeling has two main disadvantages for the magnetic field oriented {\it parallel} 
the NW axis: (i) the effective g-factors of electrons and holes cannot be probed 
separately because of the optical transitions, and (ii) this theory does not account 
for the unconventional nonlinear ZS observed. Such nonlinear features have recently 
been observed in quantum dots\cite{Jovanov2012PRB, Oberli2012PRB} and quantum wells\cite{Kotlyar2001PRB, Durnev2012PhysicaE}, 
i. e., semiconductor systems with strong quantum confinement.
 
On the other hand, the case of WZ NWs is quite different since the NWs used in 
these recent experiments have large diameter and effectively behave as a bulk 
material\cite{MishraAPL2007,FariaJunior2012JAP,Dacal2016SciRep,DeLuca2017APR} with negligible 
lateral quantum confinement. Particularly for InP WZ, it was unambiguously shown 
in the study of Tedeschi et al.\cite{Tedeschi2018} that these nonlinear features 
originate from the peculiar coupling of Landau levels (LLs) from different energy 
states in the valence band, specifically between A and B bands. Although this nonlinear 
ZS has also been observed in InGaAs\cite{DeLuca2013ACSNano} and GaAs\cite{DeLuca2017NL}, 
it remains to be shown that indeed these nonlinear features have the same origin 
and could be described in a compact analytical way. Furthermore, InAs WZ NWs have 
been investigated by recent transport experiments\cite{Das2012NatPhys,Albrecht2016Nature,Vaitiekenas2018PRL,Iorio2018arXiv}. 
However, to the best of our knowledge, no theoretical attempt has ever been made 
to compute the effective g-factors in such material.

In this paper, we analyze the ZS of novel III-V WZ materials, namely InP, InAs 
and GaAs. We provide the values for effective g-factors of different energy bands 
and highlight important contributions due to SOC effects originating from the interband 
SOC interaction. Turning to the LL physics, we apply the theoretical approach presented 
in the study of Tedeschi et al.\cite{Tedeschi2018} to show that the nonlinear ZS 
arises solely from the mixing within the valence band and it is indeed a common feature 
present in the studied materials. Based on this common mechanism responsible for the nonlinear 
features, we developed an analytical model that reliably fits the available experimental 
data, especially for InP WZ. We then extrapolate our fitted results and show that 
the nonlinear feature acts as a limiting effect to the maximum value of the ZS for 
InP. Under higher magnetic fields, such features could be observed experimentally 
in order to test the limits of our suggested model.

We organize this paper as follows: In Sec.~\ref{sec:gfactor} we discuss the effective 
g-factors calculations and the role of SOC effects. In Sec.~\ref{sec:LLs} we show 
the common nonlinear features arising in the valence band from the LL coupling. 
The effective analytical model for the nonlinear ZS is presented in Sec.~\ref{sec:modelZS} 
and we draw our conclusions in Sec.~\ref{sec:conclusions}. In the Appendix, we 
discuss the LL spectra for ZB materials.


\section{Effective g-factors and spin-orbit coupling effects}
\label{sec:gfactor}

\begin{figure}[h!]
\begin{center} 
\includegraphics{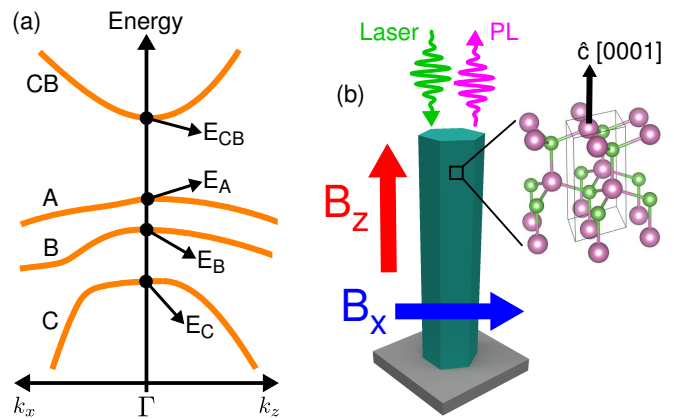}
\caption{(Color online) (a) Schematic band structure for bulk WZ crystals around 
the $\Gamma$-point indicating the different energy bands: CB, A, B and C (from top 
to bottom). The energies at $\Gamma$-point are indicated by the arrows. (b) Scheme 
of the magnetic field configurations {\it parallel} (B$_\text{z}$) and {\it perpendicular} (B$_\text{x}$) 
to the NW axis in typical magneto PL experiments. The laser excitation and the 
collected PL signal are parallel to the NW axis. The inset shows the WZ crystal 
structure with the orientation of the c-axis in [0001] direction.}
\label{fig:WZ}
\end{center}
\end{figure}

In order to evaluate the effective g-factors within the $k \cdot p$ framework, we use 
the standard perturbative approach\cite{Roth1959PR,Hermann1977PRB,LewYanVoon2009} 
that accounts for the coupling between different energy bands. 
Here we focus on the energy bands around the band gap at the $\Gamma$-point of WZ 
crystals, namely the conduction band (CB) and the top three valence bands, labeled 
A, B and C from highest to lowest energy. In Fig.~\ref{fig:WZ}(a) we depict the 
bulk WZ band structure and identify the labels for the different energy bands. 
Within this $k \cdot p$ perturbative approach, each band (two-fold degenerate) is 
described by an effective Zeeman term of the form 
\begin{equation}
H_{\text{ZS}}(B_{\alpha})=\frac{\mu_{B}}{2}B_{\alpha}g_{\alpha}\tau_{\alpha},\;\alpha=x,y,z \, ,
\end{equation}
in which the matrices $\tau_\alpha$ are the Pauli matrices for the two-fold degenerate $\Gamma$-point states 
and the effective g-factor $g_\alpha$ is obtained after evaluating
\begin{equation}
g_{\alpha}\tau^{nm}_\alpha=g_{0}\sigma^{nm}_{\alpha} -i\frac{2m_{0}}{\hbar^{2}}\underset{l\neq n,m}{\sum}\frac{\Pi_{\beta}^{nl}\Pi_{\gamma}^{lm}-\Pi_{\gamma}^{nl}\Pi_{\beta}^{lm}}{E_{n}-E_{l}}\,,
\label{eq:g}
\end{equation}
with $g_0$ ($m_0$) being the bare electron g-factor (mass), $n,m$ being the states 
of a specific energy band at $\Gamma$-point, $E_{n(l)}$ are the energy values at $\Gamma$-point and 
$\left\{ \alpha,\beta,\gamma\right\} =\left\{ x,y,z \right\}$ (or cyclic permutations). 
The matrix elements of the Pauli matrices acting on the spin 1/2 are given by $\sigma^{ab}_{\alpha} = \left\langle a \left|\sigma_{\alpha}\right| b \right\rangle$ 
and the matrix elements for the $\vec{\Pi}$ operator are given by $\Pi^{ab}_{\alpha}=\left\langle a\left|\Pi_\alpha\right|b\right\rangle$, 
with the $\vec{\Pi}$ operator written as
\begin{equation}
\vec{\Pi}=\frac{\hbar}{m_{0}}\vec{p}+\frac{\hbar^{2}}{4m_{0}^{2}c^{2}}\left[\vec{\sigma}\times\vec{\nabla}V(\vec{r})\right] \, ,
\label{eq:Pi}
\end{equation}
in which the second term describes the SOC contribution. In Fig.~\ref{fig:WZ}(b) 
we show the direction of magnetic fields with respect to the WZ crystal structure.

To compute the g-factors in Eq.~(\ref{eq:g}) we must specify a particular $k \cdot p$ Hamiltonian 
that contains the coupling among the different energy bands. For the CB, A, B and 
C bands including spin, the most general 8$\times$8 $k \cdot p$ Hamiltonian that includes 
both orbital [1st term in Eq.~(\ref{eq:Pi})] and SOC [2nd term in Eq.~(\ref{eq:Pi})] 
terms is given in Ref.~[\onlinecite{FariaJunior2016PRB}]. It is convenient 
to notice that the matrix elements $\Pi^{ab}_{\alpha}$ can be easily obtained by looking 
at the Hamiltonian terms $H_{kp}^{(1)}$ and $H_{kSO}^{(1)}$ (shown in the Appendix B 
of Ref.~[\onlinecite{FariaJunior2016PRB}]). Furthermore, an additional parameter present in the 
Hamiltonian is the k-independent SOC ($\sim\left[\vec{\nabla}V(\vec{r})\times\vec{p}\right]\cdot\vec{\sigma}$) 
between conduction and valence bands denoted by $\Delta_4$ (sometimes also called $\Delta_{sz}$\cite{Fu2008JAP}). 
The inclusion of $\Delta_4$ prevents the analytical diagonalization of the Hamiltonian 
at the $\Gamma$-point. 

The most straightforward way to calculate the g-factors is to evaluate Eq.~(\ref{eq:g}) 
numerically, especially if the Hamiltonian does not allow analytical solutions. 
However, to unambiguously identify the contribution of SOC, 
we present here the analytical expressions for CB and A band g-factors assuming 
$\Delta_4 = 0$, that allows the analytical diagonalization of the WZ Hamiltonian 
at the $\Gamma$-point (see Sec.~II.B of Ref.~[\onlinecite{Chuang1996PRB}] for instance). 
Rewriting the total g-factor of Eq.~(\ref{eq:g}) as 
$g_{\alpha}=g_{0}+\frac{2m_{0}}{\hbar^{2}}\left(L_{\alpha}+\lambda_{\alpha}\right)$ 
we can identify the orbital contributions in $L_{\alpha}$ and the SOC 
effects in $\lambda_{\alpha}$. These terms for CB and A bands read:
\begin{align}
\label{eq:gCBx}
L_{x}^{\text{CB}} & =\sqrt{2}P_{1}P_{2}ab\left(\frac{1}{\Delta_{\text{C}}}-\frac{1}{\Delta_{\text{B}}}\right) \, ,\\
\lambda_{x}^{\text{CB}} & =\frac{1}{\Delta_{\text{B}}}\bigg[2a\beta_{1}\left(a\beta_{1}-\sqrt{2}b\beta_{2}\right)\nonumber \\
 & \quad\quad\quad+bP_{1}\left(2b\beta_{2}-\sqrt{2}a\beta_{1}\right)+2a^{2}\beta_{1}P_{2}\bigg]\nonumber \\
 & +\frac{1}{\Delta_{\text{C}}}\bigg[2b\beta_{1}\left(b\beta_{1}+\sqrt{2}a\beta_{2}\right)\nonumber \\
 & \quad\quad\quad+aP_{1}\left(2a\beta_{2}+\sqrt{2}b\beta_{1}\right)+2b^{2}\beta_{1}P_{2}\bigg] \, , \nonumber
\end{align}

\begin{align}
\label{eq:gCBz}
L_{z}^{\text{CB}} & =P_{2}^{2}\left(\frac{b^{2}}{\Delta_{\text{C}}}+\frac{a^{2}}{\Delta_{\text{B}}}-\frac{1}{\Delta_{\text{A}}}\right) \, , \\
\lambda_{z}^{\text{CB}} & =\frac{1}{\Delta_{\text{A}}}\beta_{1}\left(2P_{2}-\beta_{1}\right)\nonumber \\
 & +\frac{1}{\Delta_{\text{B}}}\left[\left(a\beta_{1}-\sqrt{2}b\beta_{2}\right)^{2}+2aP_{2}\left(a\beta_{1}-\sqrt{2}b\beta_{2}\right)\right] \nonumber \\
 & +\frac{1}{\Delta_{\text{C}}}\left[\left(b\beta_{1}+\sqrt{2}a\beta_{2}\right)^{2}+2bP_{2}\left(b\beta_{1}+\sqrt{2}a\beta_{2}\right)\right]  \, , \nonumber
\end{align}

\begin{align}
\label{eq:gAz}
L_{z}^{A} & = -\frac{P_{2}^{2}}{\Delta_{\text{A}}}+2A_{7}^{2}\left(\frac{b^{2}}{\Delta_{\text{AB}}}+\frac{a^{2}}{\Delta_{\text{AC}}}\right) \, , \\
\lambda_{z}^{A} & =-\frac{1}{\Delta_{\text{A}}}\left(\beta_{1}^{2}-2\beta_{1}P_{2}\right) \nonumber \\
 & +\frac{1}{\Delta_{\text{AB}}}\left[\left(b\alpha_{1}+\sqrt{2}a\alpha_{2}\right)^{2}-2bA_{7}\left(2a\alpha_{2}+\sqrt{2}b\alpha_{1}\right)\right] \nonumber \\
 & +\frac{1}{\Delta_{\text{AC}}}\left[\left(a\alpha_{1}-\sqrt{2}b\alpha_{2}\right)^{2}+2aA_{7}\left(2b\alpha_{2}-\sqrt{2}a\alpha_{1}\right)\right] \, , \nonumber
\end{align}
with the energy differences given by $\Delta_{\text{A}}=E_{\text{CB}}-E_{\text{A}}$, 
$\Delta_{\text{B}}=E_{\text{CB}}-E_{\text{B}}$, $\Delta_{\text{C}}=E_{\text{CB}}-E_{\text{C}}$, 
$\Delta_{\text{AB}}=E_{\text{A}}-E_{\text{B}}$ and $\Delta_{\text{AC}}=E_{\text{A}}-E_{\text{C}}$. 
The values $E_{\text{CB}}$, $E_{\text{A}}$, $E_{\text{B}}$ and $E_{\text{C}}$ 
are the energies at $\Gamma$-point for the bands considered, indicated in Fig.~\ref{fig:WZ}(a). 
The $a$ and $b$ coefficients read $a = \delta/\sqrt{\delta^{2}+2(\Delta_{3})^{2}}$ 
and $b = \sqrt{2}\Delta_{3}/\sqrt{\delta^{2}+2(\Delta_{3})^{2}}$ with
$\delta=(\Delta_{1}-\Delta_{2})/2+\sqrt{(\Delta_{1}-\Delta_{2})^{2}/4+2(\Delta_{3})^{2}}$. 
The energy parameters $\Delta_1$ and $\Delta_{2,3}$ represent the crystal field splitting 
energy and the valence band SOC energies in WZ, respectively. The parameters $P_1$ 
and $P_2$ couple conduction and valence bands via the linear momentum operator, $A_7$ 
is the intra valence band coupling also mediated by $\vec{p}$, $\beta_1$ and $\beta_2$ 
are SOC terms between conduction and valence bands while $\alpha_1$ and $\alpha_2$ 
are SOC terms within the valence band only. For the precise definition of these couplings, 
please refer to the Appendix B of Ref.~[\onlinecite{FariaJunior2016PRB}].

The SOC corrections to the g-factors, shown in Eqs.~(\ref{eq:gCBx})-(\ref{eq:gAz}), 
take into account the same parameters that control the spin splitting of the energy 
bands (see for instance Eq.~(10) in Ref.[\onlinecite{FariaJunior2016PRB}] for the CB 
spin splitting parameters). Therefore, since the spin splittings of the energy bands 
are different (either in magnitude or k-dependence), so are the SOC corrections 
to the g-factors. We point out that $g_x^{\text{A}}$ is not shown simply because 
it is zero due to the symmetry of the A bands (they do not couple via the $\Pi_z$ operator 
to any other band and their different spin projections also do not couple by $\sigma_x$). 
Furthermore, we emphasize that by removing the SOC contribution (setting $\lambda_{\alpha}=0$) 
we recover the known result for the conduction band presented by Hermann and Weisbuch\cite{Hermann1977PRB}. 


\begin{table}[h!]
\caption{Calculated g-factors for the different energy bands of InP, InAs, and GaAs WZ. 
In our notation, the $z (x, y)$ direction is parallel (perpendicular) to the c-axis 
of the WZ structure, as indicated in Fig.~\ref{fig:WZ}(b). The numbers in parenthesis 
in the column with SOC indicate the approximate percentage of the SOC contribution 
to the g-factor, defined as $\left| [g(\text{SOC}) - g]/g \right|$ with $g$ being 
the g-factor without SOC effects.}
\begin{center}
{\renewcommand{\arraystretch}{1.2}
\begin{tabular*}{1.0\columnwidth}{@{\extracolsep{\fill}}
llrrrr}
\hline 
\hline 
 &  & \multicolumn{2}{c}{$\qquad$no SOC} & \multicolumn{2}{c}{$\qquad\qquad$with SOC}\tabularnewline
 & band & $g_{x}$ & $g_{z}$ & $g_{x}$ & $g_{z}$\tabularnewline
\hline 
InP$^a$ & CB &  1.72 &  1.81 &  1.29 (25) &  1.61 (11) \tabularnewline
        & A  &   0.0 & -3.30 &  0.0       & -3.05 (8)  \tabularnewline
        & B  & -3.74 &  5.35 & -3.94 (5)  &  5.12 (4)  \tabularnewline
        & C  &  5.46 &  0.24 &  5.10 (7)  &  0.47 (97) \tabularnewline
\hline 
InAs$^a$ & CB &  -5.49 &  -5.33 &  -6.82 (24)  & -6.23 (17) \tabularnewline
         & A  &   0.0  & -23.71 &   0.0        & -22.90 (3) \tabularnewline
         & B  & -19.69 & -10.81 & -19.06 (3.1) & -8.97 (17) \tabularnewline
         & C  &  14.20 &   7.57 &  14.07 (1)   &  7.70 (2)  \tabularnewline
\hline 
GaAs$^b$ & CB &  0.33 &   0.46 & \textendash{} & \textendash{} \tabularnewline
         & A  &  0.0  & -10.19 & \textendash{} & \textendash{} \tabularnewline
         & B  & -9.17 &   7.22 & \textendash{} & \textendash{} \tabularnewline
         & C  &  9.50 &  -3.43 & \textendash{} & \textendash{} \tabularnewline
\hline 
\hline
\multicolumn{4}{l}{$^{a}$$k \cdot p$ parameters from Ref.~[\onlinecite{FariaJunior2016PRB}]}\tabularnewline
\multicolumn{4}{l}{$^{b}$$k \cdot p$ parameters from Ref.~[\onlinecite{Cheiw2011PRB}]}\tabularnewline
\end{tabular*}}
\end{center}
\label{tab:gfactors}
\end{table}

Now we turn to the calculated values of the effective g-factors for InP, InAs, and GaAs WZ. For InP and 
InAs we used the parameters (which contain SOC effects) available in Ref.~[\onlinecite{FariaJunior2016PRB}] 
and for GaAs we used the parameters (without any SOC effects) from Ref.~[\onlinecite{Cheiw2011PRB}]. 
Please see footnote [\onlinecite{note:kp_models}] for a brief information on the $k \cdot p$ parameters used.
We emphasize here that the SOC effects we refer to in the $k \cdot p$ Hamiltonians are related to 
the terms that contribute to the g-factor as shown in Eqs.~(\ref{eq:gCBx})-(\ref{eq:gAz}) 
and the additional interband k-independent SOC term $\Delta_4$. Furthermore, in both 
$k \cdot p$ models\cite{FariaJunior2016PRB,Cheiw2011PRB} the usual SOC in the valence 
band is included via the parameters $\Delta_2$ and $\Delta_3$. In Table \ref{tab:gfactors} 
we show the calculated g-factors in the absence of SOC and with SOC (if the $k \cdot p$ 
model allows). As a general trend, $g_x = g_y \neq g_z$ (highlighting the anisotropy 
of the WZ structure), and the valence bands have larger g-factors (in absolute value) 
than the CB. Taking into account the SOC effects, only available for InP and InAs, 
we notice that their correction to the g-factor values are, in general, not negligible 
and with values within the typical experimental precision. For instance, the influence 
of SOC can reach contributions of $\sim 25 \%$ in CB g-factors and it is larger 
for $g_x^{\text{CB}}$ than $g_z^{\text{CB}}$. The reason for this larger SOC effect in 
$g_x^{\text{CB}}$ is due to the contribution of the parameter $P_1$ in $\lambda_x^{\text{CB}}$ 
that is absent in $\lambda_z^{\text{CB}}$ (and additionally, $P_1 > P_2$). It is 
also worth mentioning the g-factors for ZB phase. Restricting ourselves to the CB, 
the most commonly investigated case, the ZB g-factor is $g^*=1.26$ for InP, $g^*=-14.9$ 
for InAs and $g^*=-0.44$ for GaAs (with values taken from Ref.~[\onlinecite{Winkler2003}]). 
We notice that indeed the effective g-factor values are quite different for ZB and 
WZ crystal phases.

Let us now compare our theoretical g-factors values to experiments. In Table \ref{tab:gcomp} 
we compare our calculated WZ g-factors with the available magneto PL experimental 
data for InP and GaAs NWs (with diameters large enough to be considered a bulk 
system). Let us first discuss the InP case. For both $g_x^{\text{CB}}$ and 
$g_z^{\text{CB}} - g_z^{\text{A}}$, our calculated values including the SOC contributions 
provide an excellent agreement to the reported experimental values. We also point out 
the apparent inconsistency between the two experimental g-factors by showing the 
range of magnetic field used in the fitting. Since for magnetic fields along the 
z direction the ZS is nonlinear, the larger the range used in the fitting the smaller 
the g-factors will be in order to account for the sublinear features. Therefore, 
we emphasize that experimentally determined g-factors should be fitted only at the 
limit of magnetic field values where the linear regime holds. For GaAs, although 
the comparison for $g_x^{\text{CB}}$ looks reasonable, the theoretical value obtained 
for $g_z^{\text{CB}} - g_z^{\text{A}}$ is nearly twice as large as the experimental 
value. This clearly indicates that the $k \cdot p$ parameters for GaAs are not 
completely consistent and further theoretical efforts are required to build a more 
realistic model.


\begin{table}[h!]
\caption{Comparison between calculated and experimental values of the effective g-factors for InP and GaAs WZ.}
\begin{center}
{\renewcommand{\arraystretch}{1.2}
\begin{tabular*}{1\columnwidth}{@{\extracolsep{\fill}}
llrr}
\hline
\hline
 &  & $g_{x}^{\text{CB}}$ & $g_{z}^{\text{CB}}-g_{z}^{\text{A}}$\tabularnewline
\hline 
InP & expt. Ref.~[\onlinecite{Tedeschi2018}] & $1.3^{a}$ & $4.4^{a}$\tabularnewline
 & expt. Ref.~[\onlinecite{DeLuca2014NL}] &  $1.4^{b}$ & $3.4^{b}$\tabularnewline
 & this work, with SOC$^{c}$ & 1.29 & 4.66\tabularnewline
 & this work, no SOC$^{c}$ & 1.72 & 5.26\tabularnewline
\hline 
GaAs & expt. & $0.28^{d}$ & $5.4^{e}$\tabularnewline
 & this work, no SOC$^{f}$ & 0.33 & 10.83\tabularnewline
\hline 
\hline
\multicolumn{4}{l}{$^{a}$fitting up to 5 T (linear regime)}\tabularnewline
\multicolumn{4}{l}{$^{b}$fitting up to 15 T (already in nonlinear regime)}\tabularnewline
\multicolumn{4}{l}{$^{c}$$k \cdot p$ parameters from Ref.~[\onlinecite{FariaJunior2016PRB}]}\tabularnewline
\multicolumn{4}{l}{$^{d}$Ref.~[\onlinecite{Furthmeier2016NatComm}], fitting up to 0.4 T (linear regime)}\tabularnewline
\multicolumn{4}{l}{$^{e}$Ref.~[\onlinecite{DeLuca2017NL}], fitting up to 15 T (still linear regime)}\tabularnewline
\multicolumn{4}{l}{$^{f}$$k \cdot p$ parameters from Ref.~[\onlinecite{Cheiw2011PRB}]}\tabularnewline
\end{tabular*}}
\end{center}
\label{tab:gcomp}
\end{table}

Although there is, to the best of our knowledge, no magneto PL reported in pure 
InAs WZ NWs, it is important to mention that this material has recently been investigated 
in several transport experiments\cite{Das2012NatPhys,Albrecht2016Nature,Vaitiekenas2018PRL,Iorio2018arXiv}. 
Specifically, the conductance experiments by Vaitiek\ifmmode \dot{e}\else \.{e}\fi{}nas et al.\cite{Vaitiekenas2018PRL} 
using 100 nm InAs WZ NWs showed that for negative gate voltages the effective g-factor 
of CB electrons saturates to $|g^*| \sim 5$. This is significantly different from 
the bulk InAs ZB value of $g^*=-14.9$ but much closer to our predicted value for 
InAs WZ of $g_z^\text{CB}=-6.23(-5.33)$ with (without) SOC effects. Furthermore, 
the recent theoretical analysis of ZB NWs by Winkler et al.\cite{Winkler2017PRL} 
showed that orbital effects play a large role in the effective g-factors of different 
subbands. However the lowest subband still retains much of the bulk information, 
specially at 100 nm (see for instance Figs.(b),(d) in Ref.~[\onlinecite{Winkler2017PRL}]). 
Although our theoretical value for the InAs WZ g-factor seems consistent with recent 
experiments and, more importantly, it is substantially different than the ZB value, 
further investigations that account for the electrostatic environment and quantum 
confinement in WZ NWs are still required, as pointed out in the conclusions of 
Ref.~[\onlinecite{Vaitiekenas2018PRL}].


\section{Nonlinear features in Landau levels}
\label{sec:LLs}

Following the theoretical approach discussed in Ref.~[\onlinecite{Tedeschi2018}] 
to model the nonlinear ZS of InP WZ, we show in this section that indeed such 
nonlinear features are a common trend and also appear in the LLs of InAs and GaAs WZ. 
To calculate the LLs, we use the general description of an external magnetic field 
within the $k \cdot p$ framework which considers the envelope function approximation (EFA), 
combined with the minimal coupling and the Zeeman term\cite{LewYanVoon2009,Pryor2006PRL,vanBree2012PRB}. 
The mathematical procedure of the EFA leads to the general Hamiltonian
\begin{equation}
H = H_{\textrm{bulk}}\!\left[\vec{k}\rightarrow-i\vec{\nabla}+\frac{e}{\hbar}\vec{A}(\vec{r})\right]+g_{0}\frac{\mu_{B}}{2}\vec{\Sigma}\cdot\vec{B} \, ,
\label{eq:HkpBmag}
\end{equation}
in which $H_{\textrm{bulk}}(\vec{k})$ is the $k \cdot p$ bulk Hamiltonian, the replacement 
$\vec{k}\rightarrow-i\vec{\nabla}+\frac{e}{\hbar}\vec{A}(\vec{r})$ takes into account 
the EFA procedure, $g_0$ is the bare electron g-factor and $\vec{\Sigma}$ is a vector 
of the Pauli matrices describing the Zeeman term for the spins of the bulk Bloch 
functions\cite{Pryor2006PRL,vanBree2012PRB}. By choosing the vector potential with 
a single spatial dependence, we can solve the Hamiltonian (\ref{eq:HkpBmag}) numerically 
by introducing the finite differences approach\cite{Chuang1997SST}, similar to a 
quantum well treatment. The solution provides the LL spectra of the system, with 
energies denoted by $E_{\lambda}(k_{B},k_{A})$ and wave functions
\begin{equation}
\psi_{\lambda,k_{B},k_{A}}(\vec{r})=\frac{e^{i\left(k_{B}r_{B}+k_{A}r_{A}\right)}}{\sqrt{\Omega}}\sum_{l}f_{\lambda,k_{B},k_{A},l}(\rho)\,u_{l}(\vec{r}) \, ,
\label{eq:psi_sp}
\end{equation}
in which $\lambda$ is the LL label, $f_{\lambda,k_{B},k_{A},l}(\rho)$ is the envelope 
function, the summation in $l$ runs over the bulk basis states denoted by $u_{l}(\vec{r})$, 
$\Omega$ is the area of the system perpendicular to the confinement direction,
$k_B$ is the wave vector parallel to the magnetic field, $k_A$ is parallel to the 
vector potential and the spatial dependence of the vector potential is denoted by 
the coordinate $\rho$. For the two directions of magnetic field investigated here 
[indicated in Fig.~\ref{fig:WZ}(b)] we have 
$\vec{B} = B \hat{x} \Rightarrow \vec{A} = B y \hat{z}, k_B = k_x, k_A = k_z, \rho = y$ and
$\vec{B} = B \hat{z} \Rightarrow \vec{A} = B x \hat{y}, k_B = k_z, k_A = k_y, \rho = x$. 
To simplify the notation we use $\vec{B} = B \hat{x}$ as B$_\text{x}$ and $\vec{B} = B \hat{z}$ 
as B$_\text{z}$ in the remainder of the paper. We note that in experimental papers\cite{DeLuca2013ACSNano,DeLuca2014NL,DeLuca2017NL,Tedeschi2018}, 
B$_\text{x}$ and B$_\text{z}$ are typically called Voigt and Faraday configurations, 
respectively. For the numerical implementation of the Hamiltonian (\ref{eq:HkpBmag}) 
we considered the system to have a size of $L=200 \; \text{nm}$ with 401 discretization 
points (with approximately 1 point every 0.5 nm). For InP and InAs we used the 
bulk 8$\times$8 $k \cdot p$ model from Ref.~[\onlinecite{FariaJunior2016PRB}] and for GaAs 
we used the 6$\times$6 $k \cdot p$ model from Ref.~[\onlinecite{Cheiw2011PRB}].

\begin{figure}[h!]
\begin{center} 
\includegraphics{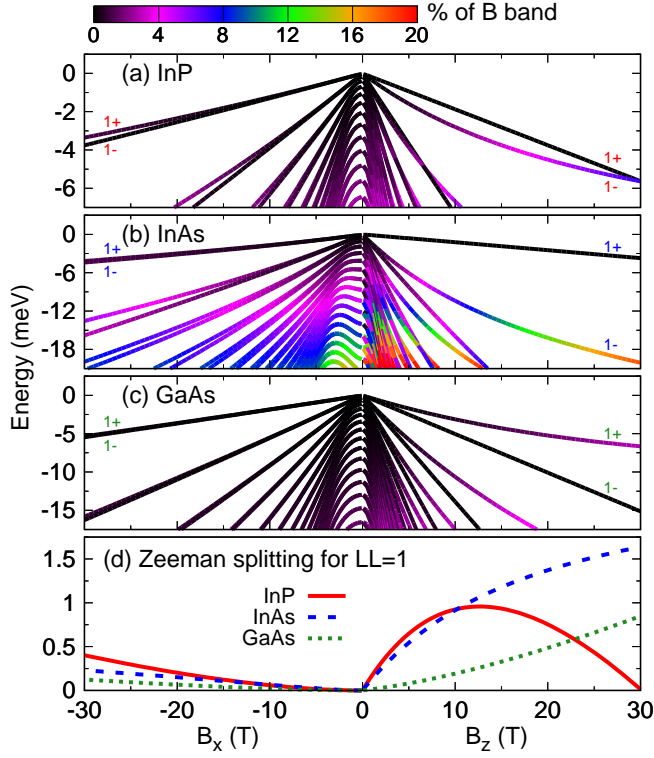}
\caption{(Color online) Calculated Landau level spectra for the A band at $k_B = k_A = 0$ 
as function of the magnetic field for (a) InP, (b) InAs and (c) GaAs. The color code 
indicates the contribution of the B valence band to the total state. The upper (lower) 
branch of the topmost Landau level is indicated by the 1+ (1-). (d) Zeeman splitting 
for the topmost Landau level. We denote with positive (negative) values in the x-axis 
the magnetic field configuration B$_\text{z}$ along z (B$_\text{x}$ along x) 
direction.}
\label{fig:LLs}
\end{center}
\end{figure}

\begin{figure}[h!]
\begin{center} 
\includegraphics{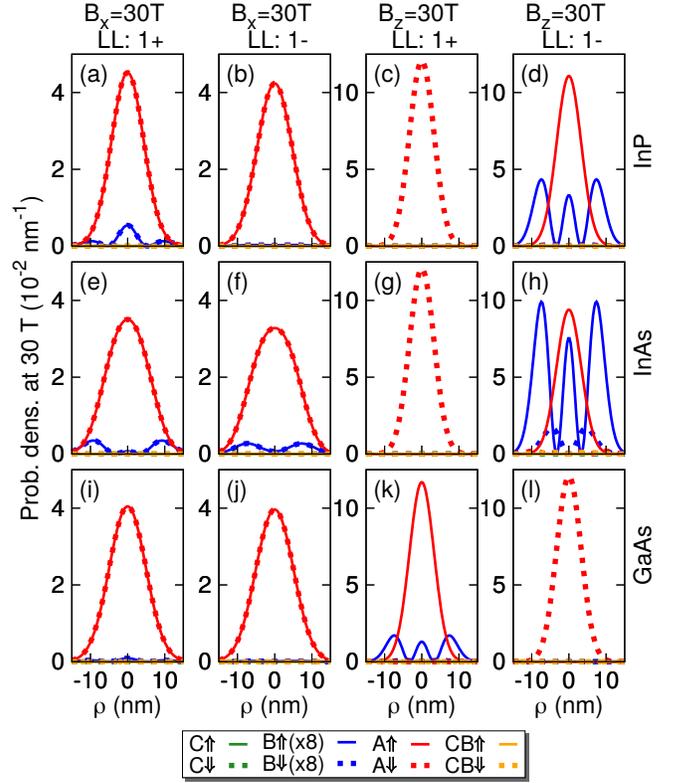}
\caption{(Color online) Probability density of the envelope functions at 30 T for 
the first LL (indicated by the labels 1+ and 1- in Fig.~\ref{fig:LLs}) along B$_\text{x}$ 
and B$_\text{z}$ for (a-d) InP, (e-h) InAs and (i-l) GaAs. The B band envelope functions 
are multiplied by a factor of 8, as indicated in the legend. The spin notation for 
the energy bands is identified with respect to the leading contribution of the bulk 
basis states\cite{note:basis}.}
\label{fig:probdens}
\end{center}
\end{figure}

Due to the coupling of the A and B bulk energy bands induced by the external magnetic 
field, the valence band LLs show a markedly nonlinear behavior\cite{Tedeschi2018}. 
This common feature can be seen in Figs.~\ref{fig:LLs}(a)-(c) for InP, InAs and GaAs, 
respectively. We notice the clear nonlinear features in the different LLs whenever 
the B band mixing is present (indicated by the color code). Although for the B$_\text{x}$ configuration this mixing 
is drastically reduced, partially because of the zero g-factor of the A band, the 
A-B mixing is indeed responsible for the slight nonzero ZS observed. As revealed 
in Ref.~[\onlinecite{Tedeschi2018}], the topmost LL (with branches indicated by 
1+ and 1- in Fig.~\ref{fig:LLs}) provides the main contribution to the excitonic 
effects and therefore the nonlinear features in the magneto PL for the B$_\text{z}$ 
configuration originate from the mixing of A and B valence bands. To highlight the 
nonlinear features, we show in Fig.~\ref{fig:LLs}(d) the ZS for the topmost LL in 
InP (solid lines), InAs (dashed lines) and GaAs (short dashed lines). We point 
out here that although the 6$\times$6 Hamiltonian for GaAs WZ\cite{Cheiw2011PRB} 
is not sufficient to describe the linear Zeeman splitting correctly, since the $P_1$ 
and $P_2$ parameters are not included, the nonlinear features in the LLs are clearly 
visible. In fact, this is an additional support to the fact that these nonlinear features 
are beyond the linear g-factor approach, which is mainly ruled by the $P_1$ and $P_2$ 
parameters as shown in Eqs.~(\ref{eq:gCBx})-(\ref{eq:gAz}).

In order to complete our analysis of the common nonlinear features of the valence 
band, we now discuss how the LL coupling manifests in the envelope functions. We 
show in Fig.~\ref{fig:probdens} the probability density of the envelope functions 
for the topmost LL (the 1+ and 1- branches shown in Fig.~\ref{fig:LLs}) in both 
magnetic field configurations at 30 T. Due to the interplay of the WZ symmetry 
and the external magnetic field, the mixing of A and B bands has a peculiar form 
that couples envelope functions with different numbers of nodes, i. e., 0 node for 
A bands and 2 nodes for B bands (further details on this coupling can be found in 
Ref.~[\onlinecite{Tedeschi2018}] and its Supplemental Material). In fact, due to 
strong SOC in InAs we also notice the contribution of B states with 1 node, see 
Figs.~\ref{fig:probdens}(f) and \ref{fig:probdens}(h). Moreover, for the B$_\text{z}$ 
configuration the coupling between envelope functions is spin dependent, which means 
that the nonlinear feature is associated to one specific type of circular polarization 
(due to the conservation of angular momentum, spins in conduction and valence band 
define the allowed transitions of circularly polarized light\cite{FariaJunior2015PRB,FariaJunior2017PRB}). 
Indeed, this is exactly the case observed in recent magneto PL experiments in InP 
WZ by Tedeschi et al.\cite{Tedeschi2018} that identified a strong nonlinear feature 
arising for a specific circular polarization of the PL spectra. On the other hand 
for the B$_\text{x}$ configuration, both spin components contribute equally and, 
as a consequence, the output light cannot be resolved in different circular polarizations.


\subsection{GaN wurtzite}

\begin{figure}[h!]
\begin{center} 
\includegraphics{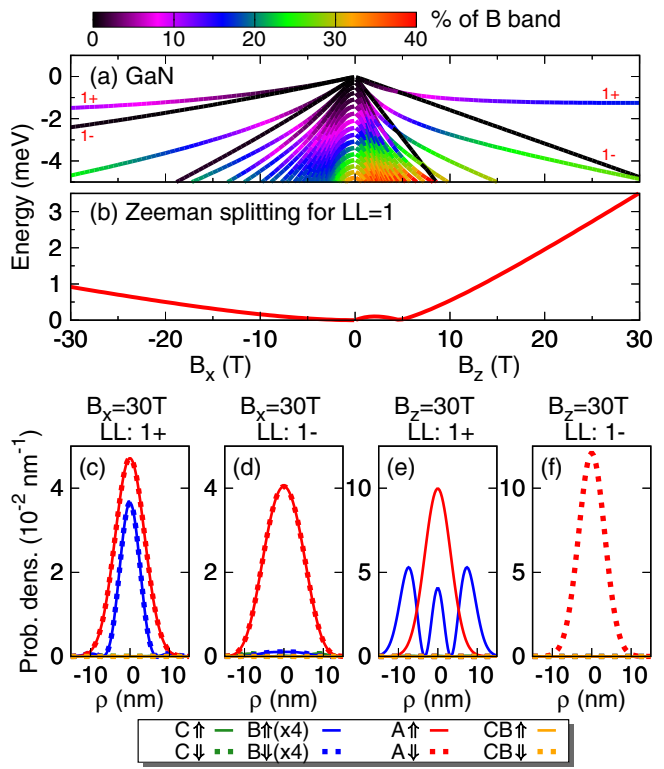}
\caption{(Color online) (a) Calculated Landau level spectra for the A band at $k_B = k_A = 0$ 
as function of the magnetic field for GaN. (b) Zeeman splitting for the topmost 
Landau level. The color code and axis notation follow Fig.~\ref{fig:LLs}. (c)-(f) 
Probability density of the envelope functions at 30 T for the first LL, indicated 
by the labels 1+ and 1- in Fig.~\ref{fig:GaN}(a), along B$_\text{x}$ and B$_\text{z}$. 
The B band envelope functions are multiplied by a factor of 4, as indicated in the legend.}
\label{fig:GaN}
\end{center}
\end{figure}

As a well established WZ compound in the family of the nitrides and recognized in 
the 2014 Nobel Prize in Physics for the efficient blue light emitting diodes\cite{Akasaki2015RMP,Amano2015RMP,Nakamura2015RMP}, 
we discuss here the case of GaN. We focus on the LL spectra of the valence bands, 
since a detailed discussion the effective g-factors in WZ GaN has been performed by 
Rodina and Meyer\cite{Rodina2001PRBb}. In Fig.~\ref{fig:GaN}(a) we show the LL spectra 
for the A band of GaN. Due to the small SOC and crystal field energies of few meV 
in GaN, the mixing of A and B bands increases in comparison to the III-V WZ materials 
discussed above (notice the color code scale in Figs.~\ref{fig:LLs} and \ref{fig:GaN}). 
For the ZS of the top most LL, shown in Fig.~\ref{fig:GaN}(b), we notice that the 
nonlinear features are present for small values of magnetic field ($<5$ T) but 
the resulting ZS is $\sim$0.1 meV, which seems to be within the experimental 
error to be properly distinguished. Therefore, the overall behavior of the ZS can 
be modeled using a linear dispersion as indicated in the experimental study of 
Rodina et al.\cite{Rodina2001PRBa}. Finally, we show in Figs.~\ref{fig:GaN}(c)-(f) 
the probability densities for the envelope functions of the LL branches 1+ and 1- 
at 30 T for the magnetic field along B$_\text{x}$ and B$_\text{z}$, and found that 
the same coupling mechanisms take place as discussed for Fig.~\ref{fig:probdens}. 

\section{Effective model for the nonlinear Zeeman splitting}
\label{sec:modelZS}

Based on the LL calculations, we showed that the physical mechanism behind the 
nonlinear ZS in the valence band is a common feature in WZ materials due to the 
mixing of A and B bands induced by magnetic field. It would be valuable to incorporate 
these features in an analytical expression that could be used to fit the experimental data 
beyond the linear ZS regime\cite{Cho1976PRB,Venghaus1977PRB}. In order to capture 
the nonlinear effects present in the LL branch with spin up, we can restrict ourselves 
to the important contributions of the A-B mixing by using the basis set 
$\left\{ f_{0,\text{A}\Uparrow}\left|\textrm{A}\Uparrow\right\rangle, f_{2,\text{B}\Uparrow}\left|\textrm{B}\Uparrow\right\rangle \right\}$, 
in which the subindices 0 and 2 refer to the number of nodes in the envelope 
functions (see Fig.~\ref{fig:probdens}). We neglect here the minor contribution of the 
envelope function with 1 node, since it appears only in InAs due to strong SOC. 
Therefore, for the coupling between A$\Uparrow$ and B$\Uparrow$ LL branches, we can write the 
following $2 \times 2$ Hamiltonian:
\begin{equation}
H=\left[\begin{array}{cc}
0 & 0\\
0 & E_{\text{B}}
\end{array}\right]+\frac{\mu_{B}}{2}B\left[\begin{array}{cc}
g_{\text{A}} & 0\\
0 & g_{\text{B}}
\end{array}\right]+B\left[\begin{array}{cc}
d_{\text{A}} & d_{\text{AB}}\\
d_{\text{AB}} & d_{\text{B}}
\end{array}\right] \, ,
\label{eq:Heff}
\end{equation}
in which the first term indicates the energy separation between A and B valence bands 
in the bulk case (we set the energy of the A band to zero), the second term is the 
ZS due to the g-factor contribution and the third term is the coupling Hamiltonian 
that mixes A and B bands, which arises from the second-order $k \cdot p$ term\cite{Chuang1996PRB,FariaJunior2016PRB}. 
Here, we assume these couplings to be parametrized by the variables $d_\text{A}$, 
$d_\text{B}$ and $d_\text{AB}$. Diagonalizing the Hamiltonian (\ref{eq:Heff}) 
we find the energy for the A$\Uparrow$ branch as
\begin{widetext}
\begin{equation}
E_{\text{A}\Uparrow}(B)=\frac{1}{2}\left[E_{\text{B}}+\frac{\mu_{B}}{2}Bg_{+}+d_{+}B+\sqrt{\left(E_{\text{B}}-\frac{\mu_{B}}{2}Bg_{-}-d_{-}B\right)^{2}+4d_{\text{AB}}^{2}B^{2}}\right] \, ,
\label{eq:Aup}
\end{equation}
with $g_{\pm}=g_{\text{A}}\pm g_{\text{B}}$ and $d_{\pm}=d_{\text{A}}\pm d_{\text{B}}$. 
For the A$\Downarrow$ branch that does not couple to any other states, unlike A$\Uparrow$, 
we have simply the linear dependence in $B$, i. e., 
\begin{equation}
E_{\text{A}\Downarrow}(B)=-\frac{\mu_{B}}{2}Bg_{\text{A}}+Bd_{\text{A}} \, ,
\label{eq:Adw}
\end{equation}
in which the first term is related to the ZS with opposite sign in the 
g-factor as compared to the A$\Uparrow$ branch and the second term is the energy 
shift of the LL branch with the same form as given in Eq.~(\ref{eq:Heff}).

In order to model the ZS obtained from the experimental PL peaks, we must take into 
account not only the ZS of the valence but also of the conduction band, since they are 
coupled via the optical transition. The total ZS can be written as the difference 
of the ZS of the conduction and valence bands, i. e., $\text{ZS}_\text{CB} - \text{ZS}_\text{A}$\cite{Cho1976PRB,Venghaus1977PRB,vanBree2012PRB}. 
For the conduction band we can use the linear g-factor ZS, $\text{ZS}_\text{CB} = \mu_BBg^{\text{CB}}_z$, and 
for the A band we use $\text{ZS}_\text{A} = E_{\text{A}\Uparrow}(B)-E_{\text{A}\Downarrow}(B)$ [shown in Eqs.~(\ref{eq:Aup}) and (\ref{eq:Adw})].
Finally, the total ZS is given by
\begin{equation}
\text{ZS}(B)=\mu_{B}B\left(g_{z}^{\text{CB}}-g_{z}^{\text{A}}\right)-\frac{1}{2}\left[\left(E_{\text{B}}-\frac{\mu_{B}}{2}Bg_{-}-d_{-}B\right)+\sqrt{\left(E_{\text{B}}-\frac{\mu_{B}}{2}Bg_{-}-d_{-}B\right)^{2}+4d_{\text{AB}}^{2}B^{2}}\right] \, ,
\label{eq:ZSBz}
\end{equation}
\end{widetext}
in which the unknown parameters are only $d_{-}$ and $\left| d_{\text{AB}} \right|$ 
if we assume the values for the effective g-factors and the energy separation of 
A and B bands given by theory or found experimentally by other means. We emphasize 
that if we set the coupling parameter $d_{\text{AB}}$ to zero in Eq.~(\ref{eq:ZSBz}), 
we recover the linear ZS already established in the literature by Refs.~[\onlinecite{Cho1976PRB,Venghaus1977PRB}]. 
Furthermore, if we set the energy separation of A and B bands to zero ($E_{\text{B}} \rightarrow 0$, 
then all the terms in Eq.~(\ref{eq:ZSBz}) become linear in the magnetic field and the nonlinearities 
vanish. This condition would be equivalent to the case of ZB crystals that have 
degenerate heavy and light hole bands at the $\Gamma$-point and therefore would present a linear 
ZS (see for instance the experimental ZS of InP ZB in the Supplemental Material 
of Ref.~[\onlinecite{Tedeschi2018}]).

\begin{figure}[h!]
\begin{center} 
\includegraphics{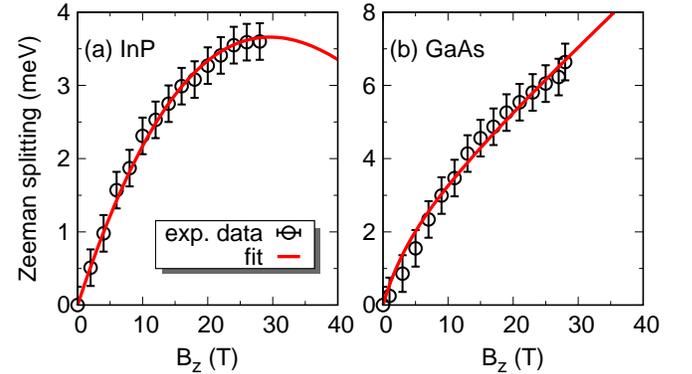}
\caption{(Color online) Comparison between the fitting of Eq.~(\ref{eq:ZSBz}) to the 
experimental Zeeman splitting for (a) InP and (b) GaAs. We use the calculated g-factors 
of Table~\ref{tab:gfactors} and the theoretical values of $E_{\text{B}} = -35.4 \; \text{meV}$ 
for InP\cite{FariaJunior2016PRB} and $E_{\text{B}} = -99.4 \; \text{meV}$ for 
GaAs\cite{Cheiw2011PRB}. The fitting procedure provides $d_{-} = 0.16 \; \text{meV/T}$, 
$|d_{\text{AB}}| = 0.43 \; \text{meV/T}$ for InP and $d_{-} = 17.63 \; \text{meV/T}$, 
$|d_{\text{AB}}| = 2.82 \; \text{meV/T}$ for GaAs. The experimental data for InP 
is taken from Ref.~[\onlinecite{Tedeschi2018}] and for GaAs from Ref.~[\onlinecite{DeLuca2017NL}]}
\label{fig:effZS}
\end{center}
\end{figure}

Applying the effective analytical ZS of Eq.~(\ref{eq:ZSBz}) to the magneto PL 
data of InP\cite{Tedeschi2018} and GaAs\cite{DeLuca2017NL}, we show in Fig.~\ref{fig:effZS} 
that this model successfully captures the experimental trends, particularly for InP WZ. 
In the fitting, we assumed the g-factors and energy separations to be known from 
theory and obtained the values for $d_{-}$ and $|d_{\text{AB}}|$ (given in the caption of Fig.~\ref{fig:effZS}). 
For GaAs we notice that the fitted parameters $d_{-}$ and $|d_{\text{AB}}|$ are nearly 
one order of magnitude larger than the values obtained for InP. We assign this 
feature to the overestimation of the GaAs g-factors in comparison with the experimental 
data in the linear regime, shown in Table~\ref{tab:gcomp}. We emphasize that 
it is beyond the scope of this study to provide reliable interband couplings 
of GaAs WZ since additional theoretical efforts are required, such as {\it ab initio} 
calculations with the correct conduction band ordering and a proper fitting of the 
$k \cdot p$ parameters, possibly including the SOC effects. Finally, we show that extrapolating 
our fitted curve up to 40 T, we observe that the ZS of InP reaches a maximum value 
and then starts to decrease. This indicates that the nonlinear features act as a 
limiting factor to the maximum ZS that can be observed. For GaAs this feature is 
not visible due to the overestimated g-factors. Therefore, additional experimental 
data at magnetic fields higher than 30 T\cite{note:bmag} could provide useful insight and also 
test the limits of the effective model presented in this study.


\section{Conclusions}
\label{sec:conclusions}

In summary, we theoretically investigated the common features of the Zeeman splitting 
in novel III-V wurtzite materials, namely InP, InAs and GaAs, using the $k \cdot p$ method. 
Specifically, we calculated the effective g-factors for the important energy bands 
around the band gap at the $\Gamma$-point (CB, A, B and C) and showed that spin-orbit 
coupling effects have appreciable contribution to the total g-factor values (contributing up to $\sim$20\% 
of CB, for instance). Our calculated values for InP and InAs g-factors are in very 
good agreement with the available experimental values. Within the Landau level picture, 
following the prescription of Tedeschi et al.\cite{Tedeschi2018}, we showed that the 
nonlinear Zeeman splitting for the B$_\text{z}$ direction is a common feature of wurtzite 
materials due to the mixing of A and B valence bands induced by the external magnetic 
field. Relying on the main mechanism behind the origin of this nonlinear feature 
allowed us to develop an effective analytical description of the Zeeman splitting 
that describes the experimental data with very good agreement, particularly for InP WZ. 
By extrapolating our fitted model, we found that the nonlinear Zeeman splitting 
of InP WZ reaches a maximum value that could be investigated experimentally under 
magnetic fields higher than 30 T. We also investigated the conventional wurtzite 
material GaN and showed that the nonlinear features are very weak to be visible 
experimentally. For zinc-blende materials, discussed in the Appendix, we showed 
that the valence band Zeeman splittings follow a strong linear behavior, specially 
for InP and GaAs.

Furthermore, our study shows that the $k \cdot p$ approach is very versatile but 
it requires reliable parameter sets for quantitative comparison with the experimental data. 
For instance, the calculated g-factors we presented for GaAs do not provide 
a good description of the experimental data, indicating that further theoretical 
efforts in extracting reliable $k \cdot p$ parameters with the correct inclusion of SOC terms are needed. 
With the ongoing interest in these novel III-V WZ materials, with recent reports 
on high-quality samples of GaP\cite{Halder2018APL} and GaSb\cite{Namazi2018AFM}, 
we believe our findings could guide future experiments and motivate further theoretical 
efforts to characterize these materials.

\begin{figure}[h!]
\begin{center} 
\includegraphics{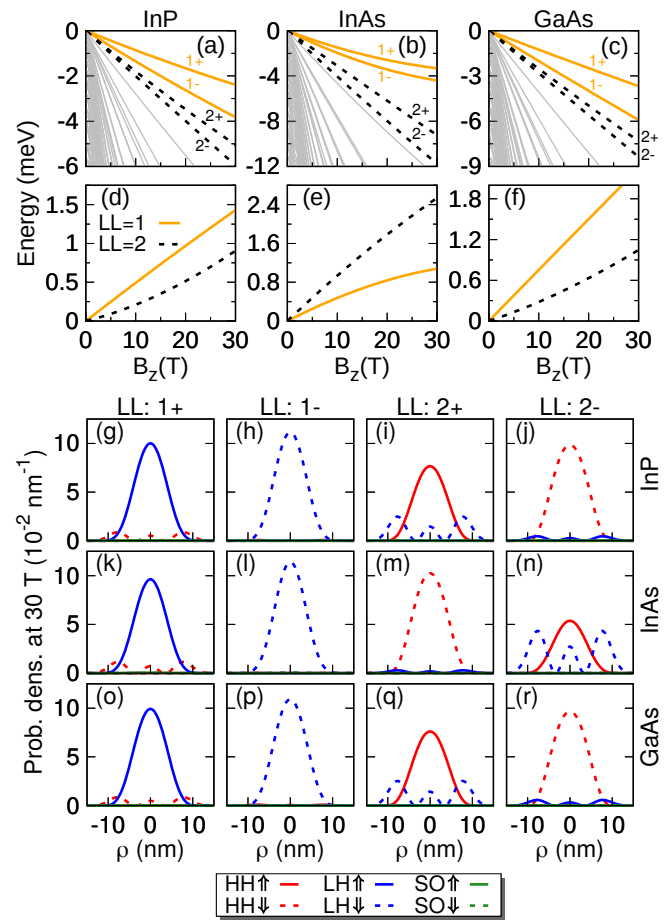}
\caption{(Color online) Calculated Landau level spectra for valence band at $k_B = k_A = 0$ 
as function of the magnetic field for (a) InP, (b) InAs and (c) GaAs with zinc 
blende structure. The upper and lower branches of the first (thick solid lines) and 
second (thick dashed lines) topmost Landau levels are indicated by the 1$\pm$ and 
2$\pm$, respectively. Zeeman splitting for the first (solid lines) and second 
(dashed lines) topmost Landau levels for (d) InP, (e) InAs and (f) GaAs. Probability 
densities at 30 T for LL=1 and LL=2 for (g)-(j) InP, (l)-(n) InAs and (o)-(r) GaAs.}
\label{fig:ZB}
\end{center}
\end{figure}


\section*{Acknowledgements}

The authors acknowledge financial support to the Alexander von Humboldt Foundation, 
Capes (grant No. 99999.000420/2016-06), SFB 1277 (B05), SFB 1170 "ToCoTronics", 
the ENB Graduate School on Topological Insulators, Awards2014 and Avvio alla 
Ricerca (Sapienza Università di Roma). P.E.F.J. is grateful to D. R. Candido, 
T. Frank, J. Lee and T. Campos for helpful dicussions.


\section*{Appendix: Zinc-blende materials}

The LL formalism discussed in Sec.~\ref{sec:LLs} can also be applied to ZB materials. 
Using the conventional Luttinger-Kohn $k \cdot p$ Hamiltonian\cite{Luttinger1955PR} 
for the valence band combined with the Zeeman term\cite{Novik2005PRB}, we calculate 
the LL spectra for InP, InAs and GaAs with ZB crystal structure assuming a magnetic 
field along the [001] axis. The effective mass parameters are taken from 
Ref.~[\onlinecite{Vurgaftman2001JAP}] and the $\kappa$ parameters from 
Ref.~[\onlinecite{Winkler2003}]. In Fig.~\ref{fig:ZB} we present our calculations 
for the LL spectra, the ZS and the probability densities focusing on the first 
and second topmost LL branches (denoted by 1$\pm$ and 2$\pm$), that have probability 
densities with majority contribution of zero nodes. Specifically, in 
Figs.~\ref{fig:ZB}(a)-(c) we show the LL spectra highlighting the LL=1 branches 
(thick solid lines) and LL=2 branches (thick dashed lines). In Figs.~\ref{fig:ZB}(d)-(f) 
we show the ZS for LL=1 and LL=2 branches. Finally, in Figs.~\ref{fig:ZB}(g)-(r) 
we show the probability densities at B = 30 T. Although in these topmost LLs in ZB 
there is also mixing of the basis states for heavy and light hole bands (degenerate 
at $\Gamma$-point), the ZS for the topmost LL is linear for InP and GaAs and slightly 
nonlinear for InAs, but not as pronounced as in WZ for the top most LLs. A small 
nonlinear ZS can also be seen for the LL=2. Finally, we point out that in typical 
magneto-PL experiments the topmost LL would be accessed via the optical transition 
and therefore the ZS for InP and GaAs would have just a linear dependence with magnetic 
field [please refer to Eq.(\ref{eq:ZSBz}) and the discussion below it for the case 
of degenerate bands with $E_{\text{B}} = 0$].



\end{document}